\documentclass[aps,prd,reprint,superscriptaddress,nofootinbib,longbibliography,floatfix]{revtex4-2}
\usepackage{booktabs}
\usepackage{amsmath,amssymb,bm,mathtools}
\usepackage{graphicx}
\usepackage{xcolor}
\usepackage{hyperref}
\usepackage{microtype}
\usepackage{color}
\hypersetup{colorlinks=true,citecolor=blue!55!black,linkcolor=blue!55!black,urlcolor=blue!55!black}
\graphicspath{{figures/}}
\newcommand{\dd}{\mathrm d}
\newcommand{\Qm}{Q_{\mathrm m}}
\newcommand{\rmn}{r_{\mathrm m}}
\newcommand{\Foff}{\mathcal F_{\mathrm{off}}}
\newcommand{\sgn}{\operatorname{sgn}}
\newcommand{\Wbar}{\overline W^{1-}}

\begin{document}

\title{Thermodynamic topology of magnetic GMGHS black holes:
the $\overline {W}^{1-}$ subclass}

\author{Dejiang Yin}
\email{yyin.dejiang@gmail.com}
\affiliation{College of Physics, Guizhou University, Guiyang 550025, People’s Republic of China}

\author{Qi-Qi liang}
\affiliation{College of Physics, Guizhou University, Guiyang 550025, People’s Republic of China}

\author{Yu-Die Wan}
\affiliation{College of Physics, Guizhou University, Guiyang 550025, People’s Republic of China}

\author{Li-yun Zhang}
\email{liy\_zhang@hotmail.com (Corresponding author)}
\affiliation{College of Physics, Guizhou University, Guiyang 550025, People’s Republic of China}

\date{\today}


\begin{abstract}
Thermodynamic topology provides a topological classification of black hole states and their branch structures in thermodynamic parameter space.
We investigate the thermodynamic topology of the charged nonextremal magnetic Gibbons--Maeda family in four-dimensional asymptotically flat Einstein--Maxwell--dilaton gravity, with particular emphasis on its string-theory member at the dilaton coupling \(a=1\), the Gibbons--Maeda--Garfinkle--Horowitz--Strominger (GMGHS) black hole. 
An analytic classification over the complete domain of regular outer horizons shows that the dilaton coupling separates the family into three limiting structures of the inverse Hawking temperature. At \(a=1\), the GMGHS defect curve approaches a nonzero lower inverse temperature endpoint and contains a single unstable branch.
In particular, the magnetic GMGHS black hole provides an explicit realization of the previously proposed \(\overline W^{1-}\) thermodynamic topological subclass within an asymptotically flat Einstein--Maxwell--dilaton solution.
This result demonstrates that the global topological number \(W\) alone does not completely characterize black hole thermodynamic topology.
The asymptotic behavior of the inverse Hawking temperature near the boundaries of the physical horizon domain provides an additional criterion for distinguishing thermodynamic topological subclasses with the same \(W\).

\end{abstract}

\maketitle

\section{Introduction}

Black hole thermodynamics combines horizon mechanics, quantum field theory, and gravitational boundary conditions \cite{1973PhRvD...7.2333B,1975CMaPh..43..199H,1977PhRvD..15.2752G}. Its phase structure is commonly characterized by thermodynamic response functions and free energies, including the Hawking--Page transition and the framework of black hole chemistry \cite{1983CMaPh..87..577H,2012JHEP...07..033K,2017CQGra..34f3001K}. 
Thermodynamic critical points have been assigned topological charges
within Duan's \(\phi\)-mapping theory
\cite{DuanGe1979,1998NuPhB.514..705D,2022PhRvD.105j4003W}.
A different construction was introduced in
Ref.~\cite{2022PhRvL.129s1101W}, where black hole equilibrium states
are represented by zeros of a vector field derived from the
generalized off-shell free energy and are characterized by their
winding numbers.

The defect construction has been studied for rotating, accelerating,
higher-curvature, NUT-charged, multicharge, lower-dimensional, and
regular black holes
\cite{2023PhRvD.107b4024W,2023PhRvD.107f4023L,
2023PhRvD.107h4002W,2023PhRvD.108h4041W,
2023PhRvD.108f6016G,2024PhLB..85638919Z,
2024EPJC...84.1294C,2023JHEP...01..102F,
2023EPJC...83..589W,2025CQGra..42l5007L,
2024PhRvD.110b4054W,2023AnPhy.45569391S}.
The dependence on matter content and thermodynamic prescription has
also been examined for charged supergravity black holes, black holes
in matter backgrounds, alternative ensembles, restricted phase space,
and generalized entropy prescriptions
\cite{2023PhRvD.107j6009G,2023EPJC...83..944R,
2024NuPhB100616649H,2024arXiv240608793W,
2024PDU....4401456G,2024EPJC...84..826T}.
Related investigations have also extended thermodynamic topology to charged dilatonic
black holes with scalar potentials
\cite{2024EPJC...84.1204H}, asymptotically AdS
Einstein--Maxwell--dilaton black holes
\cite{2026EPJC...86...78B}, and Kerr--Sen geometries
\cite{2026arXiv260324686R}. 
A classification based on the limiting behavior of the inverse
Hawking temperature \(\beta(r_h)\) near the boundaries of the physical horizon
domain and the stability of the innermost and outermost branches
was discussed in Ref.~\cite{2024PhRvD.110h1501W}. Further
classes, subclasses, and lower-dimensional realizations were identified
in gauged-supergravity, Ho\v{r}ava--Lifshitz, and BTZ black holes
\cite{2024JHEP...06..213W,2025PhRvD.111f1501W,
2025EPJC...85.1386C,2025PhLB..86539482C}. In particular,
Ref.~\cite{2025PhRvD.111f1501W} conjectured the possible
\(\overline W^{1-}\) subclass characterized by a finite lower
inverse temperature limit and a divergent large radius asymptotic limit.

Although the global topological number \(W\) provides an important
classification of black hole thermodynamic states, it does not completely characterize the boundary limiting behavior of the inverse temperature curve.
Black holes with the same value of \(W\) may exhibit
different thermodynamic evolutions due to different limiting behaviors
of \(\beta(r_h)\) at the minimal horizon radius and at the large-radius
limit. Therefore, the limiting
behavior of the inverse temperature provides an additional criterion
for distinguishing thermodynamic topological subclasses~\cite{2024PhRvD.110h1501W,2024JHEP...06..213W,
2025PhRvD.111f1501W,2025EPJC...85.1386C}.

Recent studies have further demonstrated that thermodynamic
boundaries introduced through finite cavities can modify the
topological classification by changing the endpoint contributions
to the thermodynamic degree~\cite{2026EPJC...86..929Z}. 
In contrast, the finite lower boundary considered in the present work
is determined by the admissible horizon domain of the exact
Gibbons--Maeda solution and does not originate from an external
cavity boundary.

Here we consider the asymptotically flat magnetic Gibbons--Maeda
family \cite{1988NuPhB.298..741G}. Its \(a=1\) member, corresponding to the string-theory value of the dilaton coupling, is the Gibbons--Maeda--Garfinkle--Horowitz--Strominger (GMGHS) black hole
\cite{1991PhRvD..43.3140G}, whose thermodynamic properties have been
studied in Refs.~\cite{1995PhRvD..52.4569C,2010PhRvD..81j4042W}.
Its complete nonextremal outer-horizon domain can be obtained
analytically from the exact solution, allowing us to determine the
asymptotic behavior of the inverse Hawking temperature
\(\beta(r_h)\) over the full range of the dilaton coupling \(a\).

The thermodynamic topology of the magnetic Gibbons--Maeda family
depends qualitatively on the dilaton coupling.
For \(0\leq a<1\), the inverse temperature curve has a
single minimum and two branches with local winding numbers \((+1,-1)\).
For \(a>1\), it increases monotonically from zero to infinity and
contains a single unstable branch with \(w=-1\). At \(a=1\), the
inverse temperature curve is strictly monotonic but has a finite
nonzero lower limit, with a single unstable branch. The magnetic
GMGHS black hole therefore belongs to the previously proposed
\(\Wbar\) thermodynamic topological subclass. In particular, the
\(a=1\) and \(a>1\) single defect sectors share \(W=-1\) but are
distinguished by their lower inverse temperature limits.

The paper is organized as follows. Section~\ref{sec:solution} reviews the magnetic Gibbons--Maeda solution and its thermodynamics. Section~\ref{sec:construction} introduces the thermodynamic defect construction, and Sec.~\ref{sec:classification} presents the coupling-dependent classification. Section~\ref{sec:a1} focuses on the \(a=1\) GMGHS black hole. Section~\ref{sec:discussion} discusses the physical implications, and Sec.~\ref{sec:Conclusions} summarizes the main results.

\section{Magnetic Gibbons--Maeda family}
\label{sec:solution}

\subsection{Exact solution and charge convention}

We consider the four-dimensional Einstein-frame action
\begin{equation}
 I=\frac{1}{16\pi}\int \dd^4x\sqrt{-g}\left[
 R-2(\nabla\phi)^2-e^{-2a\phi}F_{\mu\nu}F^{\mu\nu}\right],
 \label{eq:action}
\end{equation}
Here \(a\) denotes the dilaton coupling. We take \(a\geq0\) without loss of generality, since the action is invariant under the simultaneous transformation \(a\to-a\) and \(\phi\to-\phi\). In the \(r_\pm\) parametrization, the four-dimensional magnetic Gibbons--Maeda family can be written as \cite{1988NuPhB.298..741G,1991PhRvD..43.3140G}
\begin{equation}
 \Delta_\pm(r)=1-\frac{r_\pm}{r},
 \label{eq:defs}
\end{equation}
\begin{align}
 \dd s^2={}&-\Delta_+\Delta_-^{\frac{1-a^2}{1+a^2}}\dd t^2
 +\Delta_+^{-1}\Delta_-^{-\frac{1-a^2}{1+a^2}}\dd r^2\notag\\
 &+r^2\Delta_-^{\frac{2a^2}{1+a^2}}\dd\Omega_2^2,\label{eq:metric}\\
 e^{-2a(\phi-\phi_\infty)}={}&\Delta_-^{\frac{2a^2}{1+a^2}}.
 \label{eq:dilaton}
\end{align}
\begin{equation}
 F_{\theta\varphi}=\Qm\sin\theta .
 \label{eq:fields}
\end{equation}
With this normalization,
\(\Qm=(4\pi)^{-1}\int_{S^2_\infty}F\) is the magnetic charge. Electric and magnetic representatives are related by the electromagnetic duality of Einstein--Maxwell--dilaton theory, but their charge conventions differ when \(\phi_\infty\neq0\). We therefore use the magnetic charge convention throughout.

To relate our notation to Ref.~\cite{1991PhRvD..43.3140G}, Eq.~\eqref{eq:action} corresponds to its Eq.~(18), up to the overall sign convention and the normalization factor \(1/(16\pi)\), while the magnetic two-form follows from its Eq.~(8). In presenting the arbitrary-\(a\) family, Ref.~\cite{1991PhRvD..43.3140G} sets the asymptotic scalar to \(\phi_0=0\). Its Eqs.~(19)--(21) give the scalar field, the redshift function
$
\lambda^2=\Delta_+\Delta_-^{\frac{1-a^2}{1+a^2}},
$
and the areal radius
$
R=r\Delta_-^{\frac{a^2}{1+a^2}},
$
so that \(g_{\theta\theta}=R^2\) and Eqs.~\eqref{eq:metric}--\eqref{eq:dilaton} follow directly. The corresponding mass and charge relations are
$
2M=r_++\frac{1-a^2}{1+a^2}r_-,
Q_0^2=\frac{r_+r_-}{1+a^2},
$
as given in its Eqs.~(22) and (23).

The 1991 GHS paper was later corrected by an Erratum concerning the powers of \(e^{\phi_0}\) multiplying the charge~\cite{1992PhRvD..45.3888G}. For the magnetic \(a=1\) solution with nonzero \(\phi_0\), the corrected expressions contain the combination \(q^2e^{-2\phi_0}\), consistent with Ref.~\cite{2010PhRvD..81j4042W}. The shift--rescaling transformation is stated explicitly in the Erratum for \(a=1\), while its extension to arbitrary \(a\) follows from the invariance of the action.

A general finite asymptotic value \(\phi_\infty\) is restored through the shift--rescaling symmetry
$
 \phi\longrightarrow\phi+\phi_\infty, 
 \qquad
 $
 and 
 $
 F\longrightarrow e^{a\phi_\infty}F .
 \label{eq:shift}
$
This transformation leaves the combination \(e^{-2a\phi}F^2\) invariant and relates the magnetic charge at \(\phi_\infty=0\), denoted by \(Q_0\), to the magnetic charge \(\Qm\) through $\Qm=e^{a\phi_\infty}Q_0$.

Using \(Q_0^2=r_+r_-/(1+a^2)\), one then obtains
$
r_+r_-=(1+a^2)\Qm^2e^{-2a\phi_\infty}.
$
The conserved mass and magnetic charge are related to \(r_\pm\) by
\begin{equation}
 2M=r_++\frac{1-a^2}{1+a^2}r_-,
 \qquad
 r_+r_-=(1+a^2)\Qm^2e^{-2a\phi_\infty}.
 \label{eq:parameters}
\end{equation}
We identify the outer event-horizon coordinate as \(r_h\equiv r_+\) and introduce the charge-dependent length scale
\begin{equation}
 \rmn^2=(1+a^2)\Qm^2e^{-2a\phi_\infty},
 \qquad
 r_-=\frac{\rmn^2}{r_h}.
 \label{eq:rm}
\end{equation}
The Gibbons--Maeda family also includes the neutral limit \(\Qm=0\), for which the metric reduces to the Schwarzschild solution for any \(a\). In this work we focus on the nonvanishing magnetic-charge sector \(\Qm\neq0\), with \(\phi_\infty\) held fixed and finite; the choice \(\phi_\infty=0\) is included. For finite \(a\), these conditions imply \(0<\rmn<\infty\). The scale \(\rmn\) determines the lower bound of the outer-horizon coordinate within the fixed-\((\Qm,a,\phi_\infty)\) family; it is neither an independent radial parameter of a given spacetime nor an additional event horizon.

\subsection{Complete outer-horizon domain}

For fixed \((\Qm,a,\phi_\infty)\) with \(\Qm\neq0\), the regular nonextremal outer-horizon domain is
$
 r_h>r_-\quad\Longleftrightarrow\quad r_h>\rmn .
 \label{eq:domain}
$
Indeed, for \(r>r_h>r_-\), both \(\Delta_+\) and \(\Delta_-\) are positive, so that \(g_{tt}<0\), \(g^{rr}>0\), and the exterior contains no additional Killing horizon. For \(a>0\), the surface \(r=r_-\) has vanishing area and is a curvature/dilaton singularity; throughout Eq.~\eqref{eq:domain} it remains hidden inside the outer horizon. At the limiting value \(r_h=\rmn=r_-\), the singular surface coincides with the Killing horizon and the horizon area vanishes. This endpoint is therefore not a regular extremal black hole state.

All statements concerning horizon area, regularity, and thermodynamic endpoints refer to the Einstein-frame metric in Eq.~\eqref{eq:metric}. The conformally related string-frame extremal geometry is not used in the thermodynamic construction considered here. The case \(a=0\) is exceptional: \(r_-\) becomes the Reissner--Nordstr\"om inner horizon, and \(r_h=r_-\) gives the regular extremal RN black hole. Since the present analysis is restricted to nonextremal states, the equality point is excluded also for \(a=0\).

For \(a=1\), the standard extremality condition
$
\Qm^2=2M^2e^{2\phi_\infty}
$
is equivalent, using \(r_h=2M\) and \(\rmn^2=2\Qm^2e^{-2\phi_\infty}\), to \(r_h=\rmn\). The regular charged nonextremal sector is therefore characterized by
$
0<\Qm^2<2M^2e^{2\phi_\infty},
\qquad\text{or equivalently}\qquad
0<\rmn<r_h .
$

The two boundaries of the physical interval have different geometric meanings. The lower endpoint is determined by horizon regularity and is singular for \(a>0\), whereas \(r_h\to\infty\) corresponds to the asymptotic large black hole limit.

\subsection{Thermodynamic quantities}

The thermodynamic quantities follow directly from the exact solution before the off-shell free energy is introduced. From Eq.~(22) of Ref.~\cite{1991PhRvD..43.3140G} and \(r_-=\rmn^2/r_h\), the mass is obtained below. The Einstein-frame horizon area follows from Eq.~(21),
\begin{equation}
 \begin{aligned}
 A_H&=4\pi R^2(r_h)\\
 &=4\pi r_h^2
 \left(1-\frac{r_-}{r_h}\right)^{\frac{2a^2}{1+a^2}}\\
 &=4\pi r_h^2
 \left(1-\frac{\rmn^2}{r_h^2}\right)^{\frac{2a^2}{1+a^2}} .
 \end{aligned}
 \label{eq:area}
\end{equation}
Since the metric satisfies \(g_{rr}=(-g_{tt})^{-1}\), the surface gravity of the timelike Killing vector normalized at infinity is
\begin{equation}
 \kappa=\frac12\left.\frac{\dd}{\dd r}
 \left[\Delta_+\Delta_-^{\frac{1-a^2}{1+a^2}}\right]\right|_{r=r_h}
 =\frac1{2r_h}\left(1-\frac{r_-}{r_h}\right)^{\frac{1-a^2}{1+a^2}}.
 \label{eq:kappa}
\end{equation}
Using the Einstein-frame area law, with the additive constant \(S_0\) retained for comparison with Ref.~\cite{2010PhRvD..81j4042W},
$
S=\frac{A_H}{4}+S_0,
$
together with \(T_H=\kappa/(2\pi)\) and \(G=\hbar=k_B=1\), gives
\begin{align}
 M(r_h)&=\frac12\left[r_h+\frac{1-a^2}{1+a^2}\frac{\rmn^2}{r_h}\right],
 \label{eq:M}\\
 S(r_h)&=\pi r_h^2\left(1-\frac{\rmn^2}{r_h^2}\right)^{\frac{2a^2}{1+a^2}}+S_0,
 \label{eq:S}\\
 T_H(r_h)&=\frac{1}{4\pi r_h}
 \left(1-\frac{\rmn^2}{r_h^2}\right)^{\frac{1-a^2}{1+a^2}}.
 \label{eq:T}
\end{align}
The corresponding inverse Hawking temperature is defined as
$\beta(r_h)\equiv T_H^{-1}(r_h).$
Equations~\eqref{eq:M}--\eqref{eq:T}
follow from the established thermodynamic relations of the
Gibbons--Maeda solution, the Einstein-frame area law, and the surface gravity
\cite{1988NuPhB.298..741G,1991PhRvD..43.3140G,1973PhRvD...7.2333B,1993PhRvD..48.3427W,1994PhRvD..50..846I}, without using the first law as an input. Reference~\cite{2010PhRvD..81j4042W} writes the \(a=1\) entropy as \(A_H/4+S_0\) and subsequently sets \(S_0=0\). We adopt the same normalization below. Since \(S_0\) is independent of \(r_h\), it does not affect the defect equation or the local winding numbers.
Direct differentiation gives
\begin{align}
 M'&=\frac12\left[1-\frac{1-a^2}{1+a^2}\frac{\rmn^2}{r_h^2}\right],\label{eq:Mp}\\
 S'&=2\pi r_h
 \left(1-\frac{\rmn^2}{r_h^2}\right)^{\frac{a^2-1}{1+a^2}}
 \left[1-\frac{1-a^2}{1+a^2}\frac{\rmn^2}{r_h^2}\right].
 \label{eq:Sp}
\end{align}
For \(a\geq0\) and \(r_h>\rmn\), the factor in square brackets is strictly positive. Hence both \(M\) and \(S\) increase monotonically throughout the physical domain.

For completeness, the magnetic potential is
\begin{equation}
 \Psi_{\mathrm m}=\frac{\Qm e^{-2a\phi_\infty}}{r_h}.
 \label{eq:potential}
\end{equation}
At fixed \(a\) and fixed asymptotic scalar \(\phi_\infty\), Eqs.~\eqref{eq:M}--\eqref{eq:T} satisfy
\begin{equation}
 \dd M=T_H\dd S+\Psi_{\mathrm m}\dd\Qm,
 \qquad
 M=2T_H(S-S_0)+\Psi_{\mathrm m}\Qm .
 \label{eq:firstlaw}
\end{equation}
If \(\phi_\infty\) is allowed to vary, the first law acquires an additional scalar work term \cite{1996PhRvL..77.4992G,2018PhLB..782...47A}. In the present analysis, \(\phi_\infty\) is held fixed.

\section{Thermodynamic defect construction}
\label{sec:construction}

\subsection{Generalized off-shell free energy and zero-point equation}
Following the generalized off-shell formalism used in thermodynamic topology studies~\cite{2022PhRvL.129s1101W,2024PhRvD.110h1501W}, we introduce an auxiliary inverse-temperature parameter $\tau>0$ and define the off-shell free energy at fixed $(Q_m,a,\phi_\infty)$ as
\begin{equation}
 \Foff(r_h,\tau)=M(r_h)-\frac{S(r_h)}{\tau}.
 \label{eq:F}
\end{equation}
where $M$ and $S$ are the black-hole mass and entropy given in Eqs.~\ref{eq:M} and \ref{eq:S}, respectively. 
At equilibrium, $\tau=\beta(r_h)=1/T_H(r_h)$, and the off-shell free energy reduces to the on-shell free energy.

Differentiating the off-shell free energy \(\Foff=M(r_h)-S(r_h)/\tau\) with respect to \(r_h\), and using Eqs.~\eqref{eq:Mp} and \eqref{eq:Sp}, gives
\begin{equation}
 \begin{aligned}
 \frac{\partial\Foff}{\partial r_h}
 ={}&\frac12\left[1-\frac{1-a^2}{1+a^2}\frac{\rmn^2}{r_h^2}\right]\\
 &\times\left[1-\frac{4\pi r_h}{\tau}
 \left(1-\frac{\rmn^2}{r_h^2}\right)^{\frac{a^2-1}{1+a^2}}\right].
 \end{aligned}
 \label{eq:Fprime}
\end{equation}

The first factor is strictly positive throughout the physical domain \(r_h>\rmn\), and therefore introduces no additional zeros. The zero-point condition is consequently
\begin{equation}
 \tau=\beta(r_h)
 =4\pi r_h\left(1-\frac{\rmn^2}{r_h^2}\right)^{\frac{a^2-1}{1+a^2}},
 \qquad r_h>\rmn .
 \label{eq:zerocurve}
\end{equation}
As a consistency check, Eqs.~\eqref{eq:T}, \eqref{eq:Mp}, and \eqref{eq:Sp} satisfy
\begin{align}
 T_HS'&=\frac12\left[1-\frac{1-a^2}{1+a^2}\frac{\rmn^2}{r_h^2}\right]=M',\notag\\
 \partial_{r_h}\Foff&=S'\left(T_H-\frac1\tau\right).
 \label{eq:verify}
\end{align}
Thus Eq.~\eqref{eq:zerocurve} is equivalent to the equilibrium condition
\(\tau=\beta(r_h)=1/T_H(r_h)\).

\subsection{Vector field and local winding number}

For a fixed inverse temperature parameter \(\tau\), consider the domain
\begin{equation}
 D=(\rmn,\infty)\times(0,\pi)
 \label{eq:D}
\end{equation}
and define the two-component vector field
\begin{equation}
 \bm\phi=(\phi^{r_h},\phi^\Theta)
 =\left(\partial_{r_h}\Foff,-\cot\Theta\csc\Theta\right).
 \label{eq:phi}
\end{equation}
A zero of \(\bm\phi\) necessarily satisfies
$
 \Theta=\frac{\pi}{2},
 \qquad
 \tau=\beta(r_h),
$
and therefore corresponds to an equilibrium black hole state on the zero-point curve~\eqref{eq:zerocurve}.
For an isolated nondegenerate zero, the local winding number is determined by the sign of the Jacobian,
\begin{equation}
 w_i=\sgn J_i,
 \qquad
 J_i=
 \left.
 \det\frac{\partial(\phi^{r_h},\phi^\Theta)}
 {\partial(r_h,\Theta)}
 \right|_{(r_i,\pi/2)} .
 \label{eq:jacobian}
\end{equation}
Since \(\phi^{r_h}\) depends only on \(r_h\) for fixed \(\tau\), while \(\phi^\Theta\) depends only on \(\Theta\), and
$\left.\partial_\Theta\phi^\Theta\right|_{\Theta=\pi/2}=1,$
Eq.~\eqref{eq:jacobian} reduces to
\begin{equation}
 w_i
 =\sgn\!\left[
 \partial_{r_h}^2\Foff(r_i,\tau)
 \right].
 \label{eq:w}
\end{equation}
At a zero of the vector field, Eq.~\eqref{eq:verify} then gives
\begin{equation}
 w_i=\sgn T_H'(r_i)
 =-\sgn\beta'(r_i),
 \label{eq:slope}
\end{equation}
where \(S'(r_i)>0\) has been used. 
For a fixed \(\tau\) with isolated nondegenerate zeros, the total topological number is \(W=\sum_i w_i\).
Therefore, the winding number is determined by the radial derivative of the inverse temperature:
a branch with \(\beta'(r_h)<0\) has \(w=+1\), whereas a branch with \(\beta'(r_h)>0\) has \(w=-1\).
Equivalently, since the zero-point curves are plotted in the \((\beta,r_h)\) plane, the sign of \(dr_h/d\beta\) gives the same classification.

Following the terminology of the thermodynamic topology literature, black hole branches with \(w=+1\) and \(w=-1\) are referred to as thermodynamically stable and unstable, respectively. These terms characterize the local thermodynamic stability of the corresponding branches and should not be confused with dynamical stability under spacetime perturbations.

For comparison with the existing classification, Tables~\ref{tab:universal-boundary} and \ref{tab:universal-behavior} summarize the thermodynamic topological classes and subclasses relevant to the present analysis \cite{2024PhRvD.110h1501W,2025PhRvD.111f1501W,2025EPJC...85.1386C}. The boundary of the \((r_h,\Theta)\) domain is traversed counterclockwise as \(I_1\cup I_2\cup I_3\cup I_4\), with \(I_1\) located at large \(r_h\) and \(I_3\) at the lower radial boundary. DP, GP, and AP denote degenerate, generation, and annihilation points, respectively. The final row lists the proposed \(\overline W^{1-}\) subclass, whose explicit realization in the magnetic GMGHS black hole is demonstrated below.


\begin{table*}[t]
\caption{Vector-field directions on the four boundary segments and
limiting values of the inverse temperature for the thermodynamic
topological classes and subclasses relevant to the present work. Here \(r_{\min}\) denotes the lower radial boundary of the physical state space. The final column lists a representative example or proposal and, where applicable, the source of the class notation.}
\label{tab:universal-boundary}
\scriptsize
\begin{tabular*}{\textwidth}{
  @{\extracolsep{\fill}}
  lcccccccp{0.22\textwidth}
}
\hline\hline
Topological class/subclass & \(I_1\) & \(I_2\) & \(I_3\) & \(I_4\)
& \(\beta(r_{\min}^{+})\) & \(\beta(\infty)\) & \(W\)
& Representative example or proposal / classification source\\
\hline
\(W^{1-}\) & \(\leftarrow\) & \(\uparrow\) & \(\rightarrow\) & \(\downarrow\)
& \(0\) & \(\infty\) & \(-1\)
& Schwarzschild defect \cite{2022PhRvL.129s1101W}; notation \cite{2024PhRvD.110h1501W}\\

\(W^{0+}\) & \(\leftarrow\) & \(\uparrow\) & \(\leftarrow\) & \(\downarrow\)
& \(\infty\) & \(\infty\) & \(0\)
& fixed-charge RN defect \cite{2022PhRvL.129s1101W}; notation \cite{2024PhRvD.110h1501W}\\

\(W^{0-}\) & \(\rightarrow\) & \(\uparrow\) & \(\rightarrow\) & \(\downarrow\)
& \(0\) & \(0\) & \(0\)
& Schwarzschild--AdS~\cite{2022PhRvD.106f4059Y}; notation \cite{2024PhRvD.110h1501W}\\

\(W^{1+}\) & \(\rightarrow\) & \(\uparrow\) & \(\leftarrow\) & \(\downarrow\)
& \(\infty\) & \(0\) & \(+1\)
& fixed-charge RN--AdS defect \cite{2022PhRvL.129s1101W}; notation \cite{2024PhRvD.110h1501W}\\

\(W^{0-\leftrightarrow1+}\) & \(\rightarrow\) & \(\uparrow\) & \(\rightarrow/\leftarrow\) & \(\downarrow\)
& fixed temperature  & \(0\) & \(0\) or \(+1\)
& multi-charge AdS realization \cite{2024JHEP...06..213W}; notation \cite{2025PhRvD.111f1501W}\\

\(\overline W^{1+}\) & \(\rightarrow\) & \(\uparrow\) & \(\leftarrow\) & \(\downarrow\)
& fixed temperature  & \(0\) & \(+1\)
& multi-charge AdS realization \cite{2024JHEP...06..213W}; notation \cite{2025PhRvD.111f1501W}\\

\(\widehat W^{1+}\) & \(\rightarrow\) & \(\uparrow\) & \(\leftarrow\) & \(\downarrow\)
& \(\infty\) & \(0\) & \(+1\)
& dyonic AdS realization \cite{2024EPJC...84.1294C}; notation \cite{2025PhRvD.111f1501W}\\

\(\widetilde W^{1+}\) & \(\rightarrow\) & \(\uparrow\) & \(\leftarrow\) & \(\downarrow\)
& \(0\) & \(0\) & \(+1\)
& multiply rotating Kerr--AdS \cite{2025PhRvD.112l4024A}\\

\(\ddot W^{1-}\) & \(\leftarrow\) & \(\uparrow\) & \(\rightarrow\) & \(\downarrow\)
& \(0\) & \(\infty\) & \(-1\)
& hyperbolic Ho\v{r}ava--Lifshitz \cite{2025EPJC...85.1386C}\\

\(\overline W^{1-}\) & \(\leftarrow\) & \(\uparrow\) & \(\rightarrow\) & \(\downarrow\)
& fixed temperature  & \(\infty\) & \(-1\)
& predicted in Ref.~\cite{2025PhRvD.111f1501W}; GMGHS realization in this work\\
\hline\hline
\end{tabular*}
\end{table*}

\begin{table*}[t]
\caption{Thermodynamic branch structure and degenerate-point behavior of the topological classes and subclasses listed in Table~\ref{tab:universal-boundary}. ``Stable'' and ``unstable'' denote the thermodynamic stability of the corresponding black hole branches.
}
\label{tab:universal-behavior}
\scriptsize
\begin{ruledtabular}
\begin{tabular}{p{0.12\textwidth}p{0.095\textwidth}p{0.095\textwidth}p{0.19\textwidth}p{0.23\textwidth}p{0.12\textwidth}p{0.06\textwidth}}
Class/subclass & Innermost branch & Outermost branch & Low \(T\) (\(\beta\to\infty\)) & High \(T\) (\(\beta\to0\)) & DP & \(W\)\\
\hline
\(W^{1-}\) & unstable & unstable & unstable large & unstable small & in pairs & \(-1\)\\
\(W^{0+}\) & stable & unstable & unstable large + stable small & no black hole state & one more GP & \(0\)\\
\(W^{0-}\) & unstable & stable & no black hole state & unstable small + stable large & one more AP & \(0\)\\
\(W^{1+}\) & stable & stable & stable small & stable large & in pairs & \(+1\)\\
\(W^{0-\leftrightarrow1+}\) & unstable & stable & no black hole state & stable large & one more AP & \(0\) or \(+1\)\\
\(\overline W^{1+}\) & stable & stable & no black hole state & stable large & in pairs & \(+1\)\\
\(\widehat W^{1+}\) & stable & stable & unstable small + two stable small & stable large & one more GP & \(+1\)\\
\(\widetilde W^{1+}\) & unstable & stable & stable small & unstable small + stable small + stable large & one more AP & \(+1\)\\
\(\ddot W^{1-}\) & unstable & stable & unstable small & two unstable small + stable large & one more AP & \(-1\)\\
\(\overline W^{1-}\) & unstable & unstable & unstable large & no black hole state & in pairs & \(-1\)\\
\end{tabular}
\end{ruledtabular}
\end{table*}

\section{Classification by dilaton coupling}
\label{sec:classification}

The limiting behavior of the inverse Hawking temperature near the
boundaries of the physical domain follows directly from Eq.~\eqref{eq:zerocurve}:
\begin{equation}
 \lim_{r_h\to\rmn^+}\beta=
 \begin{cases}
  \infty,&0\leq a<1,\\
  4\pi\rmn,&a=1,\\
  0,&a>1,
 \end{cases}
 \qquad
 \lim_{r_h\to\infty}\beta=\infty.
 \label{eq:endpoints}
\end{equation}
The three lower endpoint limits are determined by the sign of the exponent
\((a^2-1)/(1+a^2)\), which is negative, zero, or positive for
\(a<1\), \(a=1\), or \(a>1\), respectively. The derivative of
\(\beta\) can be written as
\begin{equation}
 \frac{\dd\ln\beta}{\dd r_h}
 =\frac{1-\dfrac{3-a^2}{1+a^2}\dfrac{\rmn^2}{r_h^2}}
 {r_h\left(1-\dfrac{\rmn^2}{r_h^2}\right)}.
 \label{eq:betaprime}
\end{equation}
Since the physical domain is open at \(r_h=\rmn\), the first set of limits in Eq.~\eqref{eq:endpoints} are one-sided limits and are not values attained by regular black hole states. The images of the inverse temperature map are
\begin{equation}
 \operatorname{Im}\beta=
 \begin{cases}
  [\beta_c,\infty),&0\leq a<1,\\
  (4\pi\rmn,\infty),&a=1,\\
  (0,\infty),&a>1.
 \end{cases}
 \label{eq:betaimage}
\end{equation}

For \(0\leq a<1\), Eq.~\eqref{eq:betaprime} has a unique zero,
\begin{align}
 r_c&=\rmn\sqrt{\frac{3-a^2}{1+a^2}},\label{eq:rc}\\
 \beta_c&=4\pi\rmn\sqrt{\frac{3-a^2}{1+a^2}}
 \left[\frac{2(1-a^2)}{3-a^2}\right]^{\frac{a^2-1}{1+a^2}}.
 \label{eq:betac}
\end{align}

The point \(r_h=r_c\) is the unique minimum of \(\beta\). Consequently, no black hole defect exists for \(\tau<\beta_c\),
whereas for \(\tau>\beta_c\) there are two zeros with winding numbers \((+1,-1)\), corresponding to a stable small black hole branch and an unstable large black hole branch. At \(\tau=\beta_c\), the two 
zeros merge into an interior degenerate point with vanishing Jacobian, and the simple-zero formula~\eqref{eq:w} does not apply. For slices containing the two simple zeros, the total topological number is \(W=0\). At \(a=0\), this reduces to the familiar fixed-charge Reissner--Nordstr\"om pattern.

At \(a=1\),
$
 \beta=4\pi r_h
$
is strictly increasing on \(r_h>\rmn\), with the fixed limiting value
\(\beta(\rmn^+)=4\pi\rmn\). For \(0<\tau<4\pi\rmn\), the formal solution of the zero-point equation lies outside the physical black hole domain, and no black hole defect exists. For \(\tau>4\pi\rmn\), there is exactly one physical zero with \(w=-1\), corresponding to a single unstable black hole branch. At \(\tau=4\pi\rmn\), the zero lies on the excluded boundary \(r_h=\rmn\), so the interior winding-number formula does not apply. This endpoint structure realizes the proposed \(\overline W^{1-}\) thermodynamic topological subclass.

\begin{figure*}[t]
 \includegraphics[width=\textwidth]{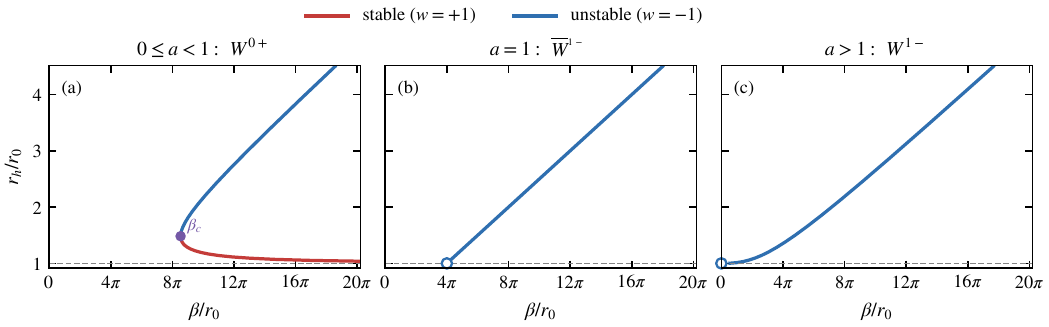}
\caption{Zero-point curves with \(\beta/r_0\) on the horizontal axis and \(r_h/r_0\) on the vertical axis. 
The reference length is chosen as \(r_0=\rmn\) for the examples shown. Red and blue denote thermodynamically stable \((w=+1)\) and unstable \((w=-1)\) black hole branches, respectively. (a) \(a=0.5\): the \(W^{0+}\) class with a single degenerate point at \(\beta_c\). 
(b) \(a=1\): the singular endpoint \((4\pi,1)\), which lies outside
the physical domain, and the \(\overline W^{1-}\) branch.
(c) \(a=1.5\): the limiting point \((0,1)\), also outside the physical domain, and the \(W^{1-}\) branch; every \(\beta>0\) intersects the curve once.}
 \label{fig:classes}
\end{figure*}

For \(a>1\), the numerator of Eq.~\eqref{eq:betaprime} is strictly positive throughout \(r_h>\rmn\). Hence \(\beta\) increases monotonically from zero to infinity, and every \(\tau>0\) admits exactly one physical zero with \(w=-1\). The limiting value \(\beta=0\) is not attained by any regular black hole, since it corresponds to the singular limit \(r_h\to\rmn^+\) and \(T_H\to\infty\). Correspondingly, \(\tau=0\) lies outside the off-shell parameter range.
Thus there is no open interval without a black hole defect for \(a>1\). By contrast, for \(a=1\) the interval \(0<\tau\leq4\pi\rmn\) contains no interior black hole defect.
These three cases exhaust the charged nonextremal magnetic Gibbons--Maeda family and are summarized in Table~\ref{tab:classification}; representative zero-point curves are shown in Fig.~\ref{fig:classes}.
\begin{table*}[t]
\caption{Thermodynamic topological classification of the charged nonextremal magnetic Gibbons--Maeda family on the physical domain \(r_h>\rmn\). 
}
\label{tab:classification}
\begin{ruledtabular}
\begin{tabular}{ccccccc}
Coupling & \(\beta(\rmn^+)\) & \(\beta(\infty)\) & Defect structure & Low \(T\) & High \(T\) & Class/subclass \\
\hline
\(0\leq a<1\)
& \(\infty\)
& \(\infty\)
& none, DP, or \((+1,-1)\)
& stable small + unstable large
& no black hole defect
& \(W^{0+}\)\\

\(a=1\)
& \(4\pi\rmn\)
& \(\infty\)
& none or \((-1)\)
& unstable large
& no black hole defect
& \(\overline W^{1-}\)\\

\(a>1\)
& \(0\)
& \(\infty\)
& \((-1)\) 
& unstable large
& unstable small as \(\tau\to0^+\)
& \(W^{1-}\)\\
\end{tabular}
\end{ruledtabular}
\end{table*}

\section{Magnetic GMGHS black hole at \(a=1\)}
\label{sec:a1}

\subsection{GMGHS solution and thermodynamics}

The value \(a=1\) corresponds to the four-dimensional low-energy string effective action \cite{1991PhRvD..43.3140G}. In this case, the magnetic Gibbons--Maeda solution reduces to the magnetic GMGHS black hole,
\begin{align}
 \dd s^2={}&-\left(1-\frac{r_h}{r}\right)\dd t^2
 +\left(1-\frac{r_h}{r}\right)^{-1}\dd r^2\notag\\
 &+r\left(r-\frac{\rmn^2}{r_h}\right)\dd\Omega_2^2,
 \label{eq:a1metric}\\
 \rmn&=\sqrt{2}\,|\Qm|e^{-\phi_\infty},
 \qquad r_h>\rmn,
 \label{eq:a1domain}\\
 M&=\frac{r_h}{2},
 \label{eq:a1M}\\
 S&=\pi\left(r_h^2-\rmn^2\right)+S_0,
 \label{eq:a1S}\\
 T_H&=\frac{1}{4\pi r_h}.
 \label{eq:a1thermo}
\end{align}
These expressions are the \(a=1\) specialization of the general magnetic Gibbons--Maeda family discussed in Sec.~\ref{sec:solution}.

In the Schwarzschild-like radial coordinate, the singular surface \(r_s=r_-\) is located at
\begin{equation}
 r_s=\frac{\Qm^2e^{-2\phi_\infty}}{M}
 =\frac{\rmn^2}{r_h}.
 \label{eq:rs}
\end{equation}
The condition that the outer horizon lie outside this singular surface,
\(r_h>r_s\), is equivalent to
$
 r_h>\rmn .
$
Thus \(\rmn\) is the lower bound of the allowed nonextremal outer-horizon radius at fixed \((\Qm,\phi_\infty)\). It is not the radial position of an additional horizon in each spacetime. The limiting configuration \(r_h\to\rmn^+\) corresponds to the singular zero-area endpoint of the Einstein-frame black hole family.

The thermodynamic quantities in Eqs.~\eqref{eq:a1M}--\eqref{eq:a1thermo} agree directly with the magnetic GMGHS thermodynamics given in Ref.~\cite{2010PhRvD..81j4042W}. Using the notation \((q,\phi_0,r_+)\), that reference gives
\begin{align}
 r_+&=2M,
 \qquad
 T_H=\frac{1}{8\pi M},
 \notag\\
 A&=4\pi r_+
 \left(
 r_+-\frac{q^2e^{-2\phi_0}}{M}
 \right),
 \notag\\
 S&=4\pi M^2
 -2\pi q^2e^{-2\phi_0}
 +S_0
 =\frac{A}{4}+S_0 .
 \label{eq:wei-dictionary}
\end{align}
With the identifications
\begin{equation}
 q=\Qm,
 \qquad
 \phi_0=\phi_\infty,
 \qquad
 r_+=r_h,
 \qquad
 \rmn^2=2\Qm^2e^{-2\phi_\infty},
\end{equation}
Eq.~\eqref{eq:wei-dictionary} reduces to
Eqs.~\eqref{eq:a1M}--\eqref{eq:a1thermo}. Following Ref.~\cite{2010PhRvD..81j4042W}, we set \(S_0=0\) in the numerical plots; the additive constant is retained in Eq.~\eqref{eq:a1S} because it does not affect any radial derivative.

The thermodynamic topology properties of this branch already follow from the general analysis in Sec.~\ref{sec:classification}. At \(a=1\),
\begin{equation}
 \beta=4\pi r_h,
 \qquad
 \tau_m=4\pi\rmn ,
\end{equation}

For this branch the radial component of the defect field reduces to
\begin{equation}
 \phi^{r_h}=\frac12-\frac{2\pi r_h}{\tau},
 \qquad
 \left.J\right|_{\bm\phi=0}=-\frac{2\pi}{\tau}<0 .
 \label{eq:a1phi}
\end{equation}
Thus the physical zero exists only for \(\tau>\tau_m\) and has
\(w=-1\).
For \(0<\tau\leq\tau_m\), no black hole defect lies in the physical domain. In the following subsection, we verify the local winding number and the corresponding global topological number directly for the magnetic GMGHS solution.

\subsection{Local winding number and global degree}

\begin{figure}[!t]
 \centering
 \includegraphics[width=\columnwidth]{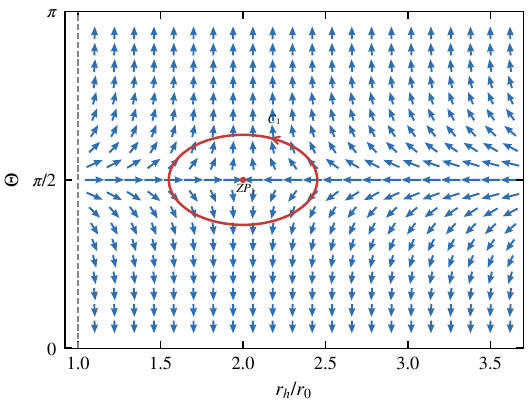}
 \caption{Normalized thermodynamic vector field for \(a=1\), \(r_0=\rmn\), and \(\tau/r_0=8\pi\). The contour \(C_1\) encloses the unique zero \(ZP_1=(2,\pi/2)\), whose winding number is \(w_1=-1\).}
 \label{fig:vector}
\end{figure}

For the representative slice \(\tau/r_0=8\pi\), with \(r_0=\rmn\), the unique zero is located at \((r_h/r_0,\Theta)=(2,\pi/2)\). Figure~\ref{fig:vector} shows the normalized thermodynamic vector field and a counterclockwise contour \(C_1\) enclosing the zero. 
For the plots, we choose the counterclockwise contour
\(r_h/r_0=2+0.45\cos\vartheta\) and
\(\Theta=\pi/2+0.42\sin\vartheta\), with
\(0\leq\vartheta\leq2\pi\).
Its image winds once clockwise around the origin, giving \(w_1=-1\).

The lower endpoint $r_h=r_m$ is excluded from the physical domain because the horizon area vanishes and the limiting spacetime geometry is singular. For $a=1$, the radial component of the thermodynamic vector field remains finite as $r_h$ approaches $r_m$ from above:
$
\lim_{r_h\to r_m^+}\phi^{r_h}
=\frac{1}{2}-\frac{2\pi r_m}{\tau}.
$
For every $\tau>\tau_m=4\pi r_m$, the vector field has a unique isolated zero in the physical domain, with $J(\phi/x)=-2\pi/\tau<0$ and winding number $w=-1$. The winding number is determined by a closed contour enclosing the zero and lying entirely within the physical domain. It remains unchanged under continuous deformations of the contour that do not cross a zero of the vector field. As $\tau$ approaches $\tau_m$ from above, the zero approaches the excluded boundary $r_h=r_m$. At $\tau=\tau_m$, the zero lies on the boundary rather than in the physical domain. Consequently, the condition that the vector field be nonzero on the boundary, which is required for the global Brouwer degree, is not satisfied at this limiting value.

\begin{figure*}[!t]
 \includegraphics[width=\textwidth]{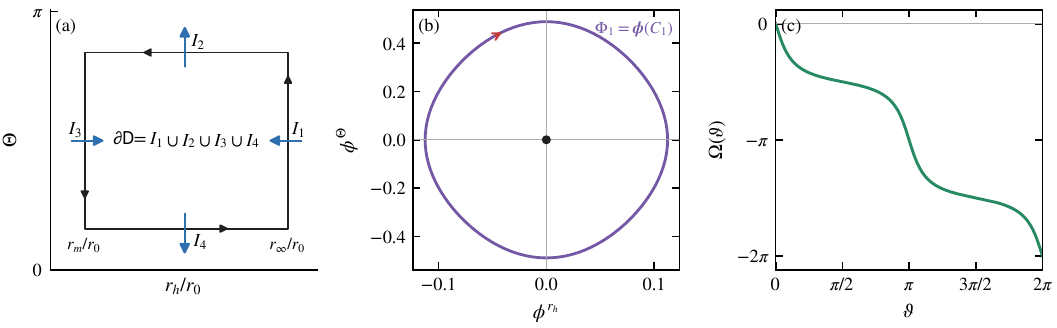}
 \caption{Local winding number and global degree for the \(a=1\) GMGHS black hole. (a) The regularized boundary \(\partial D=I_1\cup I_2\cup I_3\cup I_4\) of the thermodynamic parameter space and the asymptotic orientations of \(\boldsymbol\phi\). (b) The image \(\Phi_1=\boldsymbol\phi(C_1)\) in the \((\phi^{r_h},\phi^\Theta)\) plane of a closed contour \(C_1\) enclosing the unique defect \(ZP_1\). (c) The continuous deflection angle along \(C_1\), with \(\Delta\Omega=-2\pi\), giving \(w_1=-1\). Since the slice contains a single defect, \(W=w_1=-1\).}
 \label{fig:global}
\end{figure*}

To determine the global degree, we regularize the physical domain as
\(D_{\epsilon_r,\epsilon_\Theta,R}
=(\rmn+\epsilon_r,R)\times
(\epsilon_\Theta,\pi-\epsilon_\Theta)\), with
\(\epsilon_r>0\) and \(\epsilon_\Theta>0\), and traverse
\(\partial D_{\epsilon_r,\epsilon_\Theta,R}\) counterclockwise.
For a defect-bearing slice \(\tau>\tau_m\), \(\phi^{r_h}>0\) on the inner radial boundary and \(\phi^{r_h}<0\) on the outer radial boundary for sufficiently small \(\epsilon\) and sufficiently large \(R\), while \(\phi^\Theta\to-\infty\) as \(\Theta\to0^+\) and \(\phi^\Theta\to+\infty\) as \(\Theta\to\pi^-\). Panel (a) of Fig.~\ref{fig:global} shows these asymptotic orientations. 
Taken in counterclockwise boundary order, these orientations give one clockwise winding of the normalized vector field and hence a global degree \(W=-1\).
Panels (b) and (c) show the local contour \(C_1\): its image \(\Phi_1=\boldsymbol\phi(C_1)\) winds once clockwise around the origin and the continuous deflection angle changes by \(\Delta\Omega=-2\pi\), giving \(w_1=-1\). Since the slice contains a unique defect, \(W=\sum_iw_i=w_1=-1\).

Table~\ref{tab:match} compares the magnetic \(a=1\) GMGHS result with the defining properties of the \(\overline W^{1-}\) thermodynamic topological subclass proposed in Ref.~\cite{2025PhRvD.111f1501W}. The GMGHS branch has a fixed nonzero lower domain limit of the inverse temperature, a divergent large-radius limit, and a single unstable physical branch with \(w=-1\).

\begin{table}[t]
\caption{Comparison between the magnetic \(a=1\) GMGHS solution and the proposed \(\overline W^{1-}\) thermodynamic topological subclass.}
\label{tab:match}
\begin{ruledtabular}
\begin{tabular}{p{0.43\columnwidth}p{0.50\columnwidth}}
Criterion & GMGHS realization\\
\hline
Physical domain
& \(r_h\in(\rmn,\infty)\)\\

Inner endpoint
& \(\beta(\rmn^+)=4\pi\rmn\in(0,\infty)\)\\

Outer endpoint
& \(\beta(\infty)=\infty\)\\

Branch structure
& one increasing unstable branch\\

Local winding number
& \(w=-1\)\\

High-temperature side
& no black hole defect for \(0<\tau\leq4\pi\rmn\)\\

Low-temperature side
& one unstable large black hole\\

Global topological number
& \(W=-1\) on every defect-bearing slice\\
\end{tabular}
\end{ruledtabular}
\end{table}

\section{Discussion}
\label{sec:discussion}

For every fixed-\(\tau\) slice containing a physical GMGHS defect, the total topological number is the Brouwer degree of \(\bm\phi_\tau\) on the physical domain \(D\). At \(a=1\), the unique defect has \(w=-1\), and the local winding-number calculation agrees with the global boundary analysis, giving \(W=-1\).

The lower endpoint \(r_h=\rmn\) is not a regular black-hole
state. At \(\tau=\tau_m=4\pi\rmn\), the solution of the zero-point
equation reaches this excluded boundary, and no interior
winding number is assigned. The passage of the formal zero through
the excluded boundary at \(\tau=\tau_m\) does not represent a
thermodynamic phase transition between regular black hole states.
For \(0<\tau<\tau_m\), the formal solution lies outside the physical
domain \(r_h>\rmn\), so no black hole defect exists. This interval
therefore does not define an additional thermodynamic topological class.

For fixed nonzero magnetic charge and finite \(\phi_\infty\),
$
 \rmn=\sqrt{2}\,|\Qm|e^{-\phi_\infty}
$
is fixed by the exact GMGHS solution. At \(a=1\), the horizon area is positive only for \(r_h>\rmn\), while the singular surface satisfies \(r_s=\rmn^2/r_h<r_h\) over the same domain. As \(r_h\to\rmn^+\), the horizon area tends to zero and the singular surface approaches the outer horizon. The fixed inverse temperature endpoint is therefore approached as the regular nonextremal solutions tend to this singular limit. This differs from the regular extremal endpoint of the Reissner--Nordstr\"om family. It also differs from a black hole in a finite cavity, where the outer boundary is externally specified and the thermodynamic description involves quasilocal energy and a redshifted boundary temperature \cite{1986PhRvD..33.2092Y,1990PhRvD..42.3376B,1995PhRvD..52.4569C,2026EPJC...86..929Z}.

The defect map used here is defined at fixed \((\Qm,a,\phi_\infty)\). Holding \(\phi_\infty\) fixed is required by the chosen boundary data; allowing it to vary introduces the corresponding scalar contribution to the first law \cite{1996PhRvL..77.4992G,2018PhLB..782...47A}. Likewise, if the magnetic charge is allowed to vary, the magnetic potential enters the thermodynamic potential and the corresponding off-shell map must be modified. The topological classification obtained here therefore refers specifically to the fixed-\(\Qm\), fixed-\(\phi_\infty\) family.

The generalized off-shell free energy is used here to define the thermodynamic vector field and its zeros. This construction does not by itself establish a globally stable canonical ensemble for an asymptotically flat black hole at spatial infinity. A Euclidean canonical treatment requires appropriate boundary conditions and, for example, may be formulated by enclosing the black hole in a finite cavity \cite{1977PhRvD..15.2752G,1986PhRvD..33.2092Y,1995PhRvD..52.4569C}. The terms stable and unstable in the present analysis refer to local thermodynamic stability along the fixed-charge black hole branches and do not imply dynamical stability under spacetime perturbations.

The metric and thermodynamic quantities of the Gibbons--Maeda and GMGHS black holes are established results \cite{1988NuPhB.298..741G,1991PhRvD..43.3140G,2010PhRvD..81j4042W}. The result obtained here concerns their placement within the thermodynamic topology classification. 
The coupling dependence yields three distinct branch structures. For \(0\leq a<1\), \(\beta(r_h)\) has a single minimum \(\beta_c\), and for \(\tau>\beta_c\), the vector field has two zero points with winding numbers \(+1\) and \(-1\). As \(a\to1^{-}\), \(r_c\to r_m\) and \(\beta_c\to4\pi r_m\), with the \(w=+1\) zero point approaching the lower boundary for fixed \(\tau>4\pi r_m\). At \(a=1\), \(\beta(r_m^+)=4\pi r_m\) and a single \(w=-1\) zero point exists only for \(\tau>4\pi r_m\). For \(a>1\), \(\beta(r_m^+)=0\) and a single \(w=-1\) zero point exists for all \(\tau>0\). Thus, the \(\overline{W}^{1-}\) subclass is realized only at \(a=1\), whereas the \(a>1\) sector belongs to the standard \(W^{1-}\) class. The change in branch structure at \(a=1\) is associated with the lower boundary of the physical horizon domain, rather than an interior generation or annihilation point.

Related thermodynamic topology studies have considered electrically
charged dilatonic black holes with Liouville-type scalar potentials,
asymptotically AdS Einstein--Maxwell--dilaton solutions, and cavity
ensembles
\cite{2024EPJC...84.1204H,2026EPJC...86...78B,
2026EPJC...86..929Z}.
These constructions differ from the present asymptotically flat
Gibbons--Maeda family in their matter content, asymptotic structure,
or thermodynamic boundary conditions. In particular, the lower bound
\(r_h>\rmn\) considered here is not imposed by an external
thermodynamic boundary but follows from the intrinsic regularity
domain of the exact solution.

A comparison with the Kerr--Sen analysis of
Ref.~\cite{2026arXiv260324686R} is also relevant. Although the
nonrotating asymptotically flat limit is related to the GMGHS
geometry, the thermodynamic construction depends on the chosen
ensemble and fixed quantities. Therefore, the geometric relation
between the two solutions does not imply an identical
thermodynamic-topological structure.

The relevant distinction lies in the thermodynamic path. In the
parametrization of Ref.~\cite{2026arXiv260324686R}, the parameter
$b=q^2/(2m)$ is held fixed in the displayed off-shell construction,
whereas the present defect map is defined at fixed physical magnetic
charge $Q_m$ and fixed $\phi_\infty$. Since $b$ varies with the mass
along a fixed-charge family, the two prescriptions do not define the
same radial off-shell derivative. The present analysis instead
classifies the complete fixed-$Q_m$, fixed-$\phi_\infty$
Gibbons--Maeda outer-horizon domain and its inverse temperature
endpoint over the full dilaton-coupling range.

\section{Conclusions}
\label{sec:Conclusions}

We have conducted an analytic thermodynamic topology classification of the charged nonextremal magnetic Gibbons--Maeda family. 
Using the established thermodynamic properties of the magnetic
Gibbons--Maeda family, we construct the generalized off-shell free
energy and analyze the associated thermodynamic defect structure on
the complete physical outer-horizon domain.
The dilaton coupling separates the family into three distinct
boundary limiting behaviors of the inverse Hawking temperature. For \(0\leq a<1\), the inverse Hawking temperature has a single minimum, and the two thermodynamic defects carry winding numbers \((+1,-1)\), giving \(W=0\). For \(a>1\), the inverse temperature curve increases monotonically from zero to infinity and contains a single unstable defect with \(w=-1\). At \(a=1\), corresponding to the magnetic GMGHS black hole, the curve instead begins at the finite nonzero endpoint \(4\pi\rmn\). For \(\tau>4\pi\rmn\), there is a unique physical defect with \(w=-1\), whereas for \(0<\tau\leq4\pi\rmn\) no black hole defect exists in the physical domain. 

The magnetic GMGHS black hole therefore provides an explicit realization of the proposed \(\overline W^{1-}\) thermodynamic topological subclass in an asymptotically flat setting without an external cavity. Within the Gibbons--Maeda family, this subclass is distinguished from the standard \(W^{1-}\) class by its fixed lower inverse temperature limit, despite sharing the same global topological number \(W=-1\). This result shows that the limiting behavior of thermodynamic quantities near the boundaries of the physical state space can provide additional information for distinguishing thermodynamic topological subclasses with the same global topological number.

The present result also suggests that future searches for thermodynamic topological subclasses should take into account not only the global topological number and the number of defect branches, but also the physical origin and limiting behavior of the boundaries of the black hole state space. Exact solutions with singular, extremal, or matter-induced lower endpoints may provide useful examples for examining this issue. A systematic comparison among different thermodynamic ensembles may further clarify which endpoint properties are intrinsic to the black hole family and which arise from the imposed thermodynamic boundary conditions.

\begin{acknowledgments}
We thank the anonymous referee for the constructive comments and valuable suggestions, which have helped improve the clarity and quality of this manuscript.
This work is supported by Guizhou provincial natural science foundation project No. ZD[2026]058.
\end{acknowledgments}

\bibliography{references_scix}

\end{document}